# MCICSAM: Monte Carlo-guided Interpolation Consistency Segment Anything Model for Semi-Supervised Prostate Zone Segmentation

Guantian Huang[a], Beibei Li[b], Xiaobing Fan[c], Aritrick Chatterjee[c], Cheng Wei[d], Shouliang Qi[a], Wei Qian[a], Dianning He[e*]

[a] College of Medicine and Biological Information Engineering, Northeastern University, Shenyang 110057, China

[b] Department of Radiology, Shengjing Hospital of China Medical University, Shenyang 110004, China

[c] Department of Radiology, University of Chicago, 5841 S Maryland Ave, Chicago, IL 60637, USA

[d] School of Science and Engineering, University of Dundee, DD1 4HN, Scotland, UK

[e] School of Health Management, China Medical University, No.77 Puhe Road Shenyang North New Area, Shenyang, Liaoning, China

**Corresponding author:** Dianning He (email: hedn@cmu.edu.cn).

**ABSTRACT:**

Accurate segmentation of various regions within the prostate is pivotal for diagnosing and treating prostate-related diseases. However, the scarcity of labeled data, particularly in specialized medical fields like prostate imaging, poses a significant challenge. Segment Anything Model (SAM) is a new large model for natural image segmentation, but there are some challenges in medical imaging. In order to better utilize the powerful feature extraction capability of SAM as well as to address the problem of low data volume for medical image annotation, we use Low-Rank Adaptation (LoRA) and semi-supervised learning methods of Monte Carlo guided interpolation consistency (MCIC) to enhance the fine-tuned SAM. We propose Monte Carlo-guided Interpolation Consistency Segment Anything Model (MCICSAM) for application to semi-supervised learning based prostate region segmentation. In the unlabeled data section, MCIC performs two different interpolation transformations on the input data and incorporates Monte Carlo uncertainty analysis in the output, forcing the model to be consistent in its predictions. The consistency constraints imposed on these interpolated samples allow the model to fit the distribution of unlabeled data better, ultimately improving its performance in semi-supervised scenarios. We use Dice and Hausdorff Distance at 95th percentile (HD95) to validate model performance. MCICSAM yieldes Dice with 79.38% and 89.95%, along with improves HD95 values of 3.12 and 2.27 for transition zone and transition zone. At the same time MCICSAM demonstrates strong generalizability. This method is expected to bring new possibilities in the field of prostate image segmentation.

## 1. INTRODUCTION

Magnetic resonance (MR) imaging is widely used for the detection, localization and diagnosis of prostate cancer (PCa) [1]. Automated MR imaging segmentation of the prostate provides significant value for prostate cancer assessment, such as calculating automated PSA densities and other key imaging biomarkers [2]. Automated T2-weighted image segmentation of the transition zone (TZ) and peripheral zone (PZ) of the prostate helps to evaluate clinically significant cancers according to the PI-RADS v2.1 guideline [3], for example, PCa occurs mostly in the PZ, while benign prostatic hyperplasia occurs mainly in the TZ [4]. Therefore, precise and automatic segmentation of the prostate regions on prostate MR images is of great clinical value for the diagnosis of prostate diseases [5]. Deep learning-based methods perform well in medical image segmentation tasks due to large amounts of high-quality labeled data for training [6], and only experts can provide reliable and accurate labeled data [7]. Therefore, especially in the field of prostate-specific medical images, labeled data are scarce and difficult to obtain, making it difficult to segment the prostate region.

Recently, Segmentation Anything Model (SAM) [8] has gained great attention in the direction of semantic segmentation of natural images because of its powerful generalization ability. Although it performs well on natural images, recent studies have also shown that SAM performs poorly on medical image segmentation [9, 10]. Large-scale computer vision models usually determine the boundaries between different segmented regions based on differences in pixel intensities, which is effective in natural images [11]. However, because of the complexity and similarity of the internal regional organization of the prostate on MR [12], large-scale computer vision models are not directly applicable to prostate zone segmentation.

The SAM input prompt guides the model in outputting the final result. This is also the reason why it is not effective to use segment everything directly when processing images in some specialized fields. When segmentation with prompt is performed, SAM is actually implemented is a binary classification segmentation task. The model is based on the features of a selected point, and the target object where this point is located is segmented from the background. Ma et al. [10] have achieved significant success in medical image segmentation by adapting SAM to MedSAM with medical images. However, MedSAM has some difficulties in segmenting prostate applications due to the complexity of similar structures in the internal regions of the prostate and the impracticality of using cues for downward guided segmentation of large amounts of unlabeled data.

Based on the small number of annotations in medical images and the need for accurate segmentation, the integration of semi-supervised learning into SAM, which is powerful in feature extraction, is expected to bring better results in the field of prostate segmentation. Recently, a team from Fudan University has already proposed SemiSAM [13] based on this semi-supervised learning idea. However, SemiSAM only uses the main network framework for the points of unlabeled data prediction region as prompt to pass into the SAM after the loss calculation. This learning method is better for the overall tissue differentiation, but it is difficult to achieve the segmentation of the prostate zones.

The recent model SMAed [14] using Low-Rank Adaptation (LoRA) [15] fine-tuned SAM was proposed. SAMed does not require the guidance of prompt for multi-organ segmentation of abdominal images. In order to investigate the segmentation effect of SAM framework for semi-supervised learning of prostate, we use SAMed as the backbone of semi-supervised learning. We used the idea of semi-supervised learning to incorporate into the powerful feature extraction capability of SAM, and evaluate the effect of the SAM framework in medical image data with insufficient amount of labeling.

The main contributions of our proposed work are summarized as follows:

• This study presented a Monte Carlo-guided interpolation consistency-based (MCIC) framework for segmenting 2D MR images of the prostate region. The specific generalization of the SAMed framework was improved by having the same distribution among pairs of unlabeled data. We proposed Monte Carlo-guided Interpolation Consistency Segment Anything Model (MCICSAM) for application to semi-supervised learning based prostate region segmentation.

• The segmentation accuracy was improved by calculating the consistency loss function after adding Monte Carlo uncertainty analysis to interpolation consistency training (ICT) [16].

• With the integration of existing semi-supervised learning methods into SAMed, the feasibility and superiority of LoRA fine-tuned SAM in the field of prostate segmentation were comprehensively evaluated to enhance segmentation accuracy and robustness.

## 2. Proposed methodology

### 2.1. Backbone architecture

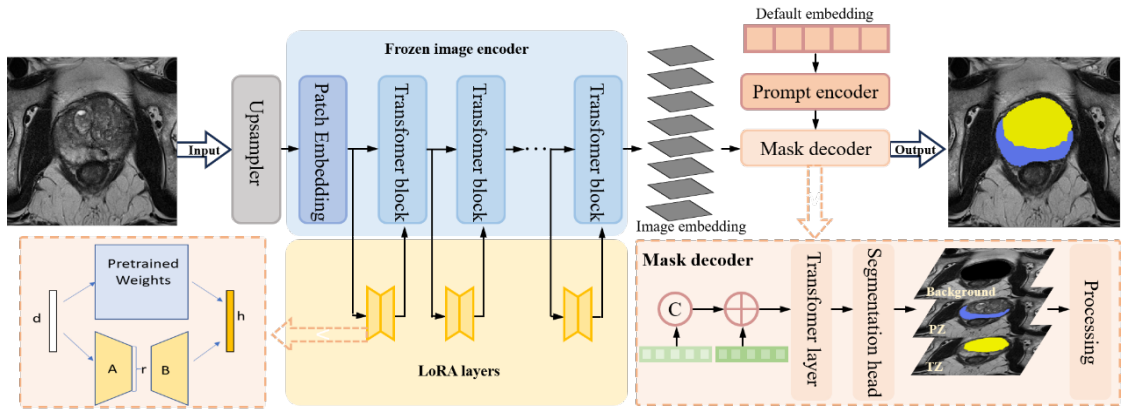

**Fig 1.** SAMed framework

The framework of SAMed is illustrated in Figure 1. LoRA allows the model to learn information more specifically adapted to the new task by introducing a downscaling and then upscaling branch next to the original model. During training, the parameters of SAM are fixed and only the downscaling matrix A and the upscaling matrix B are trained, and the output is superimposed with the parameters of the original model. Compared with fine-tuning all the parameters in the SAM, LoRA allows the SAM to update a small portion of the parameters during the training process of medical images, which ensures the segmentation performance and reduces the difficulty of deploying and storing the fine-tuned model.

SAMed freezes all parameters in the image encoder and designs a trainable LoRA layer for each Transformer module. SAMed uses a default prompt so that no prompt information is needed to perform automatic segmentation during inference. The image encoder is frozen and additional trainable LoRA layers are inserted into SAM for medical image feature extraction. The mask decoder in SAM consists of a Transformer layer and a segmentation header. SAMed modifies SAM's segmentation header to customize the output for each category. Unlike the vague prediction of SAM, SAMed predicts each category in a deterministic way.

*2.2. MCIC semi-supervised learning*

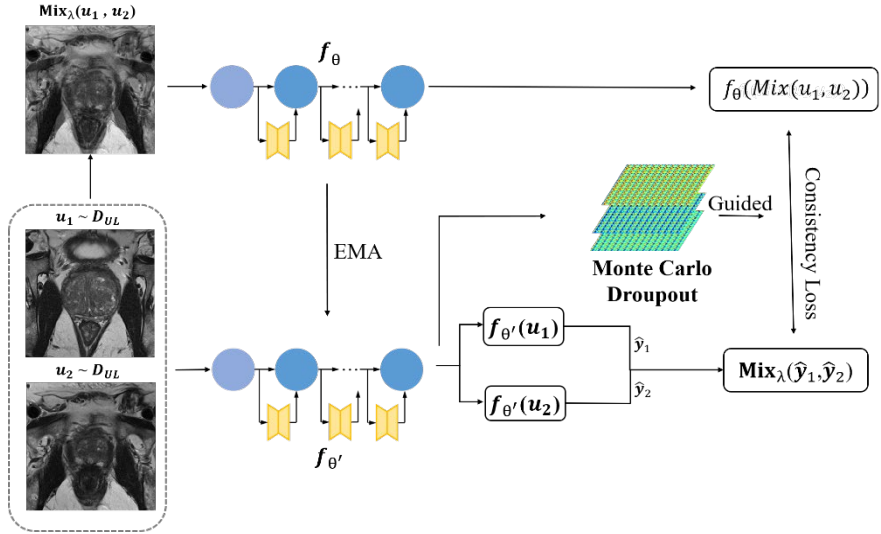

**Fig 2.** MCIC framework

Figure 2 illustrates the framework of MCIC, which divides the model into a student model $f_\theta$ and a teacher model $f_{\theta'}$. Teacher model uses it to generate learning goals for students, and the student model uses the goals provided by the teacher for learning. The weight of the teacher model is obtained from the weighted average of the student model's time memory. In the context of the MCIC, the Exponential Moving Average (EMA) mechanism is employed for parameter updates. The updated formula is expressed as:

$$\theta'_t = \alpha \theta'_{t-1} + (1-\alpha)\theta_t \tag{1}$$

where α signifies momentum, $\theta'_t$ is the teacher network and $\theta_t$ is the student network. For instance, when α is set to 0.99, the teacher network retains 99 % of its parameters unchanged during each update, incorporating 1 % from the student network.

For unlabeled data, we split the unlabeled data into two parts as $u_1$ and $u_2$. we hope that the model will be consistent in its prediction by interpolating the two inputs. The consistency constraints imposed on these interpolated samples allow the model to better fit the distribution of unlabeled data, which ultimately improves its performance in semi-supervised learning scenarios.

The formula for the interpolation operation on unlabeled data awakening is:

$$Mix_\lambda(u_1, u_2) = \lambda u_1 + (1-\lambda)u_2 \tag{2}$$

where two inputs a and b are linearly interpolated based on a mixing coefficient λ.

MCIC trains the student model $f_\theta$ to provide consistent predictions at interpolations of unlabeled images:

$$f_\theta(Mix(u_1, u_2)) \approx Mix_\lambda(f_{\theta'}(u_1), f_{\theta'}(u_2)) \tag{3}$$

where $f_\theta(Mix(u_1, u_2))$ is the prediction of the student model $f_\theta$ on the mixup of unlabeled images $u_1$ and $u_2$, $Mix_\lambda(f_{\theta'}(u_1), f_{\theta'}(u_2))$ is the mixup of predictions generated by the teacher model $f_{\theta'}$ on the same unlabeled images.

The ICT framework is based on two-dimensional image classification. However, segmentation on medical images such as prostate is not like in natural image classification, where there is a strong uncertainty in certain regions of the segmentation result. Therefore, we add an uncertainty-aware analysis so that the student model can gradually learn from more reliable targets. The teacher model not only generates target predictions, but also estimates the uncertainty of each target. The student model is optimized for more accurate segmentation results guided by the estimated uncertainty by making the most use of the consistency loss.

The prediction equation for the teacher model is:

$$\hat{y}_t = Mix_\lambda(\frac{1}{T}\sum_{t=1}^{T} f_{\theta'_t}(u_1), \frac{1}{T}\sum_{t=1}^{T} f_{\theta'_t}(u_2)) \qquad (4)$$

We use Monte Carlo Dropout to estimate uncertainty. We perform a random dropout on each input data and then perform T random forward passes on the teacher model. For a given input data, predictions are made by varying randomness and the probability distribution of each prediction is calculated using the softmax function. Then, the average of these probability distributions is used as the result and interpolated to get the teacher model to the output result $\hat{y}_t$. This helps to identify which predictions are unreliable and thus improves the credibility of the model and makes it more accurate in the predicting process.

*2.3. Loss functions of supervised and semi-supervised learning*

SAMed adopts cross entropy (CE) loss and Dice loss to supervise the finetuning process. The supervised learning loss function can be described as:

$$L_{supervised} = \lambda_1 CE(\hat{y}_l, D(u_l)) + \lambda_2 Dice(\hat{y}_l, D(u_l)) \qquad (5)$$

where $\hat{y}_l$ presents labeled data predictions, $u_l$ presents labels of the labeled data, $D$ denotes as the downsample operation to make $u_l$ the same as $\hat{y}_l$. $\lambda_1$ and $\lambda_2$ represent the loss weights to balance the influence between these two loss terms.

The loss function for semi-supervised learning uses consistency loss. The consistency loss function formula is as follows:

$$L_{Con} = \frac{1}{N}\Sigma_{i=1}^{N}(\hat{y}_\theta - \hat{y}_{\theta'})^2 \qquad (6)$$

where $\hat{y}_\theta$ and $\hat{y}_{\theta'}$ represents the predictions of the student model and the teacher model.

The Loss of the entire model consists of the $L_{supervised}$ and $w(t) \times L_{con}$. The weight $w(t)$ is incrementally increased after each iteration. This incremental increases in $w(t)$ amplifies the significance of the consistency regularization loss, aiding the model in effectively capturing and maintaining consistency in its predictions.

**3. Experimental setup**

*3.1. Datasets*

For the prostate region segmentation, we used the ProstateX [17-19] dataset, which provided publicly available ground truth annotations introduced by Meyer et al. [20]. The dataset contained multisite prostate MR scans of healthy individuals, patients with cancer, and patients with hyperplasia under a variety of conditions. The dataset contained 346 T2w axial volumes. Of these, 98 volumes were associated with labels for the PZ and TZ. Importantly, 248 masses were unlabeled to facilitate our semi-supervised learning strategy. To ensure methodological consistency, we retained 20 labeled samples for testing purposes. We also used the Medical Segmentation Decathlon (MSD) [21] prostate dataset, ISBI [19, 22] dataset and our private dataset collected from Shengjing Hospital of China Medical University in China to verify the robustness of segmentation.

These 3D volumes were cropped into 2D images with a fixed size of 256 × 256 pixels to normalize the input dimensions and reduced possible noise or irrelevant information. The intensity values were normalized by z-score standardization method [23]. This reduced sensitivity to changes in the distribution of the input data, which helped the model learn features more efficiently and updated weights more stably during the training process.

*3.2. Evaluation metrics*

We used Dice [24] and Hausdorff Distance [25] at 95th percentile (HD95) to evaluate these experiments. The Dice equation is as follows:

$$Dice = \frac{2 \times TP}{2 \times TP + FP + FN} \tag{7}$$

where FN, FP, TP and TN are false negative, false positive, true positive and true negative, respectively.

The HD95 equation is as follows:

$$\text{HD95}(M, N) = \max \left\{ \text{percentile}_{95} \left( \min_{n \in N} ||m - n|| \right)_{m \in M}, \text{percentile}_{95} \left( \min_{m \in M} ||m - n|| \right)_{n \in N} \right\} \tag{8}$$

where M and N represent the prediction mask and the ground truth, respectively. It is calculated based on the 95th percentile of the distance between the boundary points of M and N. The purpose of using this metric for evaluation is to eliminate the effect of outliers.

*3.3. Model configuration*

All experiments were conducted on the deep learning framework PyTorch and parallelized on a single RTX4090 GPU. We empirically set the EMA decay to 99. The batch sizes for both labeled and unlabeled data were set to 32. We set the learning rate to 0.005 and used the AdamW optimizer [26]. We ramped up consistency loss component weight $w$ during the first 100 epochs using a Gaussian ramp-up curve $\exp(-4(1 - T)^2)$, where T advanced linearly from zero to one during the ramp-up period. This ensured that the target loss was dominated by the supervised loss term at the outset. This prevented the network from falling into a degenerate solution and failing to make meaningful target predictions for unlabeled data. For uncertainty estimation, we set $T = 12$ to balance the quality of uncertainty estimation and training efficiency.

**4. Results and discussion**

To evaluate the proposed scheme, we compare the supervised learning framework with some of the more commonly used segmentation models, including SAM, U-Net [27] and Swin Transformer [28]. Table 1 quantitatively shows the segmentation results of the prostate region on the ProstateX dataset using different supervised learning methods. Figure 3 demonstrates the results of supervised learning segmentation. It can be seen that SAM is less effective in prostate image segmentation, indicating that SAM cannot be directly applied to medical image segmentation. The SAM with LoRA fine-tuning shows strong performance, with Dice reaching 72.25% in the PZ region, 85.50% in the TZ region, and 3.30 in the HD95 of the TZ region. SAMed is only slightly worse than U-Net in HD95 metrics in the PZ region, so we use SAMed as a backbone to test the performance of the SAM framework in semi-supervised learning.

**Table 1**

Results of Dice and HD95 for four segmentation networks segmented under supervised learning effects.

|                  | PZ       |       | TZ       |       |
|------------------|----------|-------|----------|-------|
|                  | Dice (%) | HD95  | Dice (%) | HD95  |
| U-Net            | 71.26,   | **3.67** | 83.76 | 3.47  |
| Swin Transformer | 64.44    | 6.78  | 82.39    | 4.91  |
| SAM              | 40.07    | 40.00 | 49.51    | 45.65 |
| SAMed            | **72.25** | 4.93 | **85.50** | **3.30** |

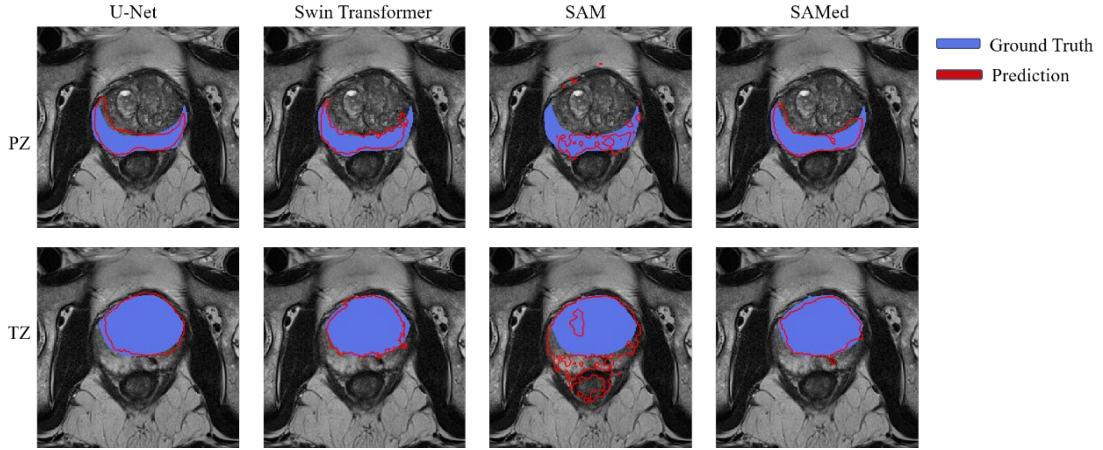

**Fig 3.** Results of four segmentation networks segmented under supervised learning on ProstateX dataset.

To test our method MCICSAM, we collect commonly used semi-supervised learning methods including mean teacher [29] (MT), uncertainly-aware mean teacher [30] (UAMT) and ICT as a comparative experiment to compare their performance on three backbones. The performance of the commonly used popular methods for semi-supervised learning is compared with our method under different backbone as demonstrated by Table 2. Figure 4 demonstrates the segmentation results of four semi-supervised learning methods combined with SAMed as backbone on prostate images. The results show that SAMed exhibits significant advantages when used as a backbone for semi-supervised learning segmentation of prostate medical images. Compared to U-Net and Swin Transformer, semi-supervised learning methods can bring higher performance in the SAMed framework. Our method achieves the best performance for both Dice and HD95 in both PZ and TZ regions. Our method achieves 79.38% Dice and 3.12 HD95 in the PZ region, and 89.95% Dice and 2.27 HD95 in the TZ region. Through this kind of comparative experiments, we are able to clearly see the influence of different backbones on the semi-supervised learning effect, which further validates the potential and superiority of the SAM framework in practical applications. We can clearly see the effect of different backbone on the semi-supervised learning effect, which further validates the potential and superiority of SAM framework in practical applications.

**Table 2**

Semi-supervised learning combining different backbone and MCICSAM methods for segmentation of Dice with HD95 results.

| Backbone | Semi-supervised methods | PZ Dice (%) | PZ HD95 | TZ Dice (%) | TZ HD95 |
|---|---|---|---|---|---|
| U-Net | MT | 75.22 | 4.49 | 87.19 | 5.86 |
| | UAMT | 75.42 | 4.54 | 87.20 | 6.00 |
| | ICT | 75.19 | 3.27 | 86.60 | 4.65 |
| Swin Transformer | MT | 70.11 | 4.27 | 84.23 | 3.87 |
| | UAMT | 67.14 | 5.44 | 83.28 | 4.59 |
| | ICT | 66.52 | 5.62 | 83.20 | 4.81 |
| SAMed | MT | 78.21 | 4.80 | 87.96 | 3.61 |
| | UAMT | 79.06 | 4.37 | 88.39 | 3.00 |
| | ICT | 79.22 | 3.71 | 89.08 | 2.76 |
| | MCICSAM | **79.38** | **3.12** | **89.95** | **2.27** |

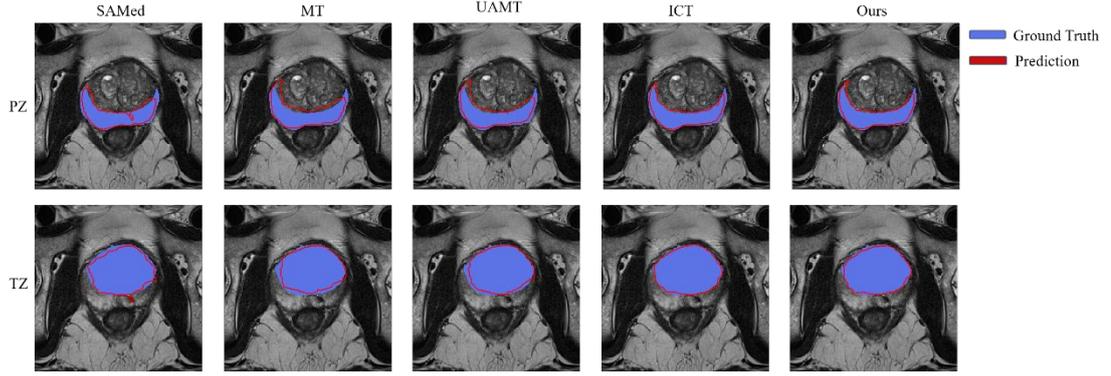

**Fig 4.** SAMed supervised learning and segmentation results on ProstateX dataset combining multiple semi-supervised learning.

In order to evaluate the effect of the number of labeled data on the experimental results, we design ablation experiments with different numbers of patients and selected 58, 38, 18 and 8 labeled patients for training, while the number of test patients as well as the number of unlabeled patients were kept constant. Table 3 shows the segmentation results for supervised learning with different number of patients, the semi-supervised learning framework ICT before the improvement and our proposed uncertainty-guided MCIC. The results exhibit that our method accuracy has higher performance most of the time. It can also be seen that the decrease in the number of patients has less impact on the SAMed framework combined with semi-supervised learning when the number of markers is between 18 and 58 individuals. This setup allows us to systematically analyze and compare the impact of different amounts of labeled data on the model performance. By verifying the importance of the amount of data on the model performance, we observe the segmentation performance of the model with reduced labeled data and the effectiveness of the semi-supervised learning approach when the amount of data is low. We observe the segmentation performance of the model in the presence of reduced labeled data and the effectiveness of semi-supervised learning methods when the amount of data is low by verifying the importance of the amount of data on the performance of the model.

**Table 3**

Results of SAMed as backbone semi-supervised learning versus supervised learning for different number of annotations available.

| # Patient | Methods | PZ | | TZ | |
|---|---|---|---|---|---|
| | | Dice (%) | HD95 | Dice (%) | HD95 |
| 8 | supervised | 52.41 | 17.08 | 68.57 | 16.33 |
| | ICT | 69.20 | 5.87 | 82.30 | **6.28** |
| | Ours | **70.35** | **5.60** | **82.84** | 8.12 |
| 18 | supervised | 56.93 | 14.63 | 72.49 | 12.93 |
| | ICT | 76.92 | 4.29 | 86.80 | 3.59 |
| | Ours | **77.81** | **4.09** | **87.23** | **3.18** |
| 38 | supervised | 66.18 | 6.78 | 80.73 | 8.98 |
| | ICT | 78.58 | 3.92 | 88.47 | 2.89 |
| | Ours | **78.82** | **3.89** | **88.68** | **2.71** |
| 58 | supervised | 72.25 | 4.93 | 85.50 | 3.30 |
| | ICT | 79.22 | 3.71 | 89.08 | 2.76 |
| | Ours | **79.38** | **3.12** | **89.95** | **2.27** |

In order to test the robustness of the semi-supervised learning model, we conducted comparative experiments on the MSD, ISBI, and our own datasets. Firstly, we trained the model directly on the datasets without using the pre-trained weights on the ProstateX dataset to verify the effect of pre-training on the model performance. Second, we trained the model on the ProstateX dataset to obtain the trained weights including SAMed supervised learning and SAMed combined with MCIC, and then directly applied these weights to the MSD, ISBI, and our own datasets for testing to evaluate the performance of the model on different datasets. Finally, after training the model on the ProstateX dataset, we used the MSD, ISBI, and our own datasets for fine-tuning training to understand the model performance improvement after fine-tuning on the new datasets.

Table 4 shows the quantitative results of segmentation on the other datasets. Figure 5 shows the segmentation results of these comparison experiments on the MSD dataset. With these four experiments, we systematically evaluate the performance difference of semi-supervised learning models in different situations and verify their robustness. It is seen that MCIC is more adaptable under the SAMed framework, and then replacing it with a new dataset only requires fine-tuning the training to get good results.

**Table 4**

Results of Dice and HD95 for transfer learning segmentation on the MSD dataset, ISBI dataset and our dataset.

|  |  | PZ | | TZ | |
|---|---|---|---|---|---|
|  |  | Dice (%) | HD95 | Dice (%) | HD95 |
| MSD | SAMed training | 46.77 | 11.87 | 68.33 | 6.37 |
|  | SAMed-pretrained | 41.86 | 26.35 | 70.93 | 13.21 |
|  | Ours-pretrained | 58.28 | 7.69 | 81.72 | 5.67 |
|  | Ours-pretrained fine-tuning | **61.18** | **5.83** | **82.57** | **4.72** |
| ISBI | SAMed training | 52.22 | 8.98 | 73.20 | 7.19 |
|  | SAMed-pretrained | 66.19 | 8.94 | 80.20 | 7.79 |
|  | Ours-pretrained | 72.35 | 7.55 | 82.74 | 3.61 |
|  | Ours-pretrained fine-tuning | **77.33** | **4.08** | **84.03** | **4. 50** |
| Our dataset | SAMed training | 76.16 | 3.48 | 73.82 | 6.61 |
|  | SAMed-pretrained | 41.22 | 15.67 | 43.43 | 26.61 |
|  | Ours-pretrained | 49.78 | 9.09 | 58.81 | 14.25 |
|  | Ours-pretrained fine-tuning | **81.10** | **2.26** | **78.52** | **4.05** |

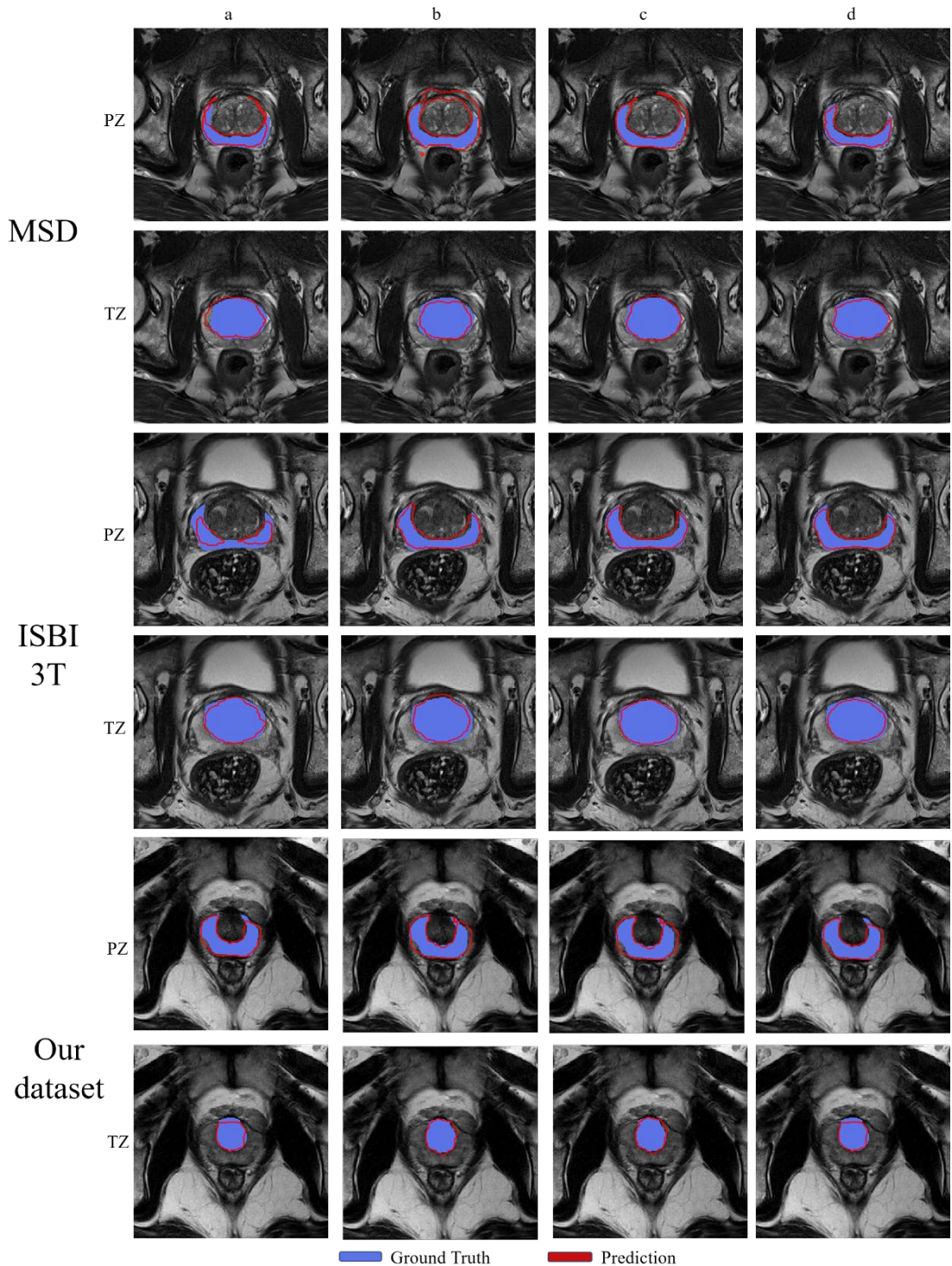

**Fig 5.** Results of segmentation on the MSD, ISBI and our datasets. (a) SAMed supervised learning trained on the datasets followed by segmentation. (b) SAMed supervised learning trained on ProstateX dataset with weights applied to the datasets for segmentation. (c) Our proposed method trained on ProstateX dataset with weights applied to the datasets for segmentation. (d) Weights of our proposed method trained on the ProstateX dataset are applied to the datasets for segmentation after training fine-tuning on the datasets.

## 5. Conclusion

In this study, we propose an improved semi-supervised learning segmentation model for T2 prostate region in the SAM framework, which incorporates an uncertainty-aware semi-supervised learning method on the ICT model for segmenting different regions inside the prostate from 2D MR images. We explore the uncertainty of the model to improve the quality of the objective. The effectiveness of MCICSAM and its strong generalization ability are verified by comparison with other semi-supervised methods. Future work includes studying the impact of different uncertainty estimation methods and applying our framework to other semi-supervised learning medical image segmentation problems.

**CRediT authorship contribution statement**

**Guantian Huang:** Writing – original draft, Methodology, Data curation, Visualization, Validation. **Beibei Li:** Data curation, Formal analysis, Investigation, Methodology. **Shouliang Qi:** Conceptualization, Resources. **Wei Qian:** Conceptualization, Resources. **Dianning He:** Writing – review & editing, Methodology, Supervision, Project administration.

**Declaration of competing interest**

The authors declare that they have no known competing financial interests or personal relationships that might influence the work reported in this paper.

**Data availability**

Data are available upon request. Please contact corresponding author at hedn@bmie.neu.edu.cn.

**Acknowledgments**

This work was partly supported by the National Natural Science Foundation of China under Grant 82001781, the Science and Technology Foundation of Liaoning Provincial under Grant 2023 MSBA-096, and the Fundamental Research Funds for the Central Universities under Grant N2419003.